\documentclass[12pt,a4paper,english,aps,pra,preprint,groupedaddress,showpacs,showkeys]{revtex4-1}
\usepackage[utf8]{inputenc}
\usepackage{color}
\usepackage{babel}
\usepackage{units}
\usepackage{bm}
\usepackage{amsmath}
\usepackage{amssymb}
\usepackage{graphicx}
\usepackage[unicode=true,
 bookmarks=true,bookmarksnumbered=false,bookmarksopen=true,bookmarksopenlevel=2,
 breaklinks=false,pdfborder={0 0 0},backref=false,colorlinks=true]
 {hyperref}
\hypersetup{pdftitle={Angular distributions in J/ψ→ppπ(η) decays},
 pdfauthor={S. G. Salnikov},
 citecolor=blue,urlcolor=blue,linkcolor=blue}

\makeatletter

\pdfpageheight\paperheight
\pdfpagewidth\paperwidth

\usepackage[justification=centerlast,labelsep=period,font=small,figurename=Fig.]{caption}

\makeatother

\begin{document}

\title{Angular distributions in $J/\psi\to p\bar{p}\pi^{0}(\eta)$ decays}

\author{V. F. Dmitriev}

\email{V.F.Dmitriev@inp.nsk.su}

\selectlanguage{english}%

\author{A. I. Milstein}

\email{A.I.Milstein@inp.nsk.su}

\selectlanguage{english}%

\author{S. G. Salnikov}

\email{S.G.Salnikov@inp.nsk.su}

\selectlanguage{english}%

\affiliation{\textit{Budker Institute of Nuclear Physics, 630090, Novosibirsk,
Russia}}

\affiliation{\textit{Novosibirsk State University, 630090, Novosibirsk, Russia}}

\date{\today}
\begin{abstract}
The differential decay rates of the processes $J/\psi\to p\bar{p}\pi^{0}$
and $J/\psi\to p\bar{p}\eta$ close to the $p\bar{p}$ threshold are
calculated with the help of the $N\bar{N}$ optical potential. The
same calculations are made for the decays of $\psi(2S)$. We use the
potential which has been suggested to fit the cross sections of $N\bar{N}$
scattering together with $N\bar{N}$ and six pion production in $e^{+}e^{-}$
annihilation close to the $p\bar{p}$ threshold. The $p\bar{p}$ invariant
mass spectra is in agreement with the available experimental data.
The anisotropy of the angular distributions, which appears due to
the tensor forces in the $N\bar{N}$ interaction, is predicted close
to the $p\bar{p}$ threshold. This anisotropy is large enough to be
investigated experimentally. Such measurements would allow one to
check the accuracy of the model of $N\bar{N}$ interaction.
\end{abstract}
\maketitle
\noindent \global\long\def\re#1{\qopname\relax{no}{Re}#1}

\section{Introduction}

The cross section of the process \mbox{$e^{+}e^{-}\to p\bar{p}$}
reveals an enhancement near the threshold~\mbox{\cite{Aubert2006,Lees2013,Achasov2014,Akhmetshin2015}}.
The enhancement near the $p\bar{p}$ threshold has been also observed
in the decays \mbox{$J/\psi\to\gamma p\bar{p}$}, \mbox{$B^{+}\to K^{+}p\bar{p}$},
and \mbox{$B^{0}\to D^{0}p\bar{p}$}~\mbox{\cite{Bai2003,Abe2002,Abe2002a}}.
These observations led to numerous speculations about a new resonance~\cite{Bai2003},
$p\bar{p}$ bound state~\mbox{\cite{Datta2003,Ding2005,Yan2005}}
or even a glueball state~\mbox{\cite{Kochelev2006,Li2006,He2007}}
with the mass near two proton mass. This enhancement could appear
due to the nucleon-antinucleon final-state interaction. It has been
shown that the behavior of the cross sections of $N\bar{N}$ production
in $e^{+}e^{-}$ annihilation can be explained with the help of Jülich
model~\mbox{\cite{Haidenbauer2006a,Haidenbauer2014}} or slightly
modified Paris model~\mbox{\cite{dmitriev2007final,dmitriev2014isoscalar}}.
These models also describe the energy dependence of the proton electromagnetic
form factors ratio $\left|G_{E}^{p}/G_{M}^{p}\right|$. A strong dependence
of the ratio on the energy close to the $p\bar{p}$ threshold is a
consequence of the tensor part of the $N\bar{N}$ interaction.

Another phenomenon has been observed in the process of $e^{+}e^{-}$
annihilation to mesons. A sharp dip in the cross section of the process
\mbox{$e^{+}e^{-}\to6\pi$} has been found in the vicinity of the
$N\bar{N}$ threshold~\mbox{\cite{Aubert2006a,Aubert2007b,Akhmetshin2013,Lukin2015,Obrazovsky2014}}.
This feature is related to the virtual $N\bar{N}$ pair production
with subsequent annihilation to mesons~\mbox{\cite{Haidenbauer2015,Dmitriev2016}}.
In Ref.~\cite{Dmitriev2016} a potential model has been proposed
to fit simultaneously the cross sections of $N\bar{N}$ scattering
and $N\bar{N}$ production in $e^{+}e^{-}$ annihilation. This model
describes the cross section of the process \mbox{$e^{+}e^{-}\to6\pi$}
near the $N\bar{N}$ threshold as well. A qualitative description
of this process was also achieved using the Jülich model~\cite{Haidenbauer2015}.

In this paper we investigate the decays \mbox{$J/\psi\to p\bar{p}\pi^{0}$}
and \mbox{$J/\psi\to p\bar{p}\eta$} taking the $p\bar{p}$ final-state
interaction into account. Investigation of these processes has been
performed in Ref.~\cite{Kang2015} using the chiral model. However,
the tensor part of the $p\bar{p}$ interaction was neglected in that
paper. To describe the $p\bar{p}$ interaction we use the potential
model proposed in Ref.~\cite{Dmitriev2016}, where the tensor forces
play an important role. The account for the tensor interaction allows
us to analyze the angular distributions in the decays of $J/\psi$
and $\psi(2S)$ to $p\bar{p}\pi^{0}(\eta)$ near the $p\bar{p}$ threshold.
The parameter of anisotropy is large enough to be studied in the experiments.

\section{Decay amplitude}

Possible states for a $p\bar{p}$ pair in the decays \mbox{$J/\psi\to p\bar{p}\pi^{0}$}
and \mbox{$J/\psi\to p\bar{p}\eta$} have quantum numbers $J^{PC}=1^{--}$
and $J^{PC}=1^{+-}$. The dominating mechanism of the $p\bar{p}$
pair creation is the following. The $p\bar{p}$ pair is created at
small distances in the $^{3}S_{1}$ state and acquires an admixture
of $^{3}D_{1}$ partial wave at large distances due to the tensor
forces in the nucleon-antinucleon interaction. The $p\bar{p}$ pairs
have different isospins for the two final states under consideration
($I=1$ for the $p\bar{p}\pi^{0}$ state, and $I=0$ for the $p\bar{p}\eta$
state), that allows one to analyze two isospin states independently.
Therefore, these decays are easier to investigate theoretically than
the process $e^{+}e^{-}\to p\bar{p}$, where the $p\bar{p}$ pair
is a mixture of different isospin states.

We derive the formulas for the decay rate of the process \mbox{$J/\psi\to p\bar{p}x$},
where $x$~is one of the pseudoscalar mesons $\pi^{0}$ or $\eta$.
The following kinematics is considered: $\bm{k}$~and $\varepsilon_{k}$~are
the momentum and the energy of the $x$ meson in the $J/\psi$ rest
frame, $\bm{p}$~is the proton momentum in the $p\bar{p}$ center-of-mass
frame, $M$~is the invariant mass of the $p\bar{p}$ system. The
following relations hold: 
\begin{align}
 & p=\left|\bm{p}\right|=\sqrt{\frac{M^{2}}{4}-m_{p}^{2}}\,, &  & k=\left|\bm{k}\right|=\sqrt{\varepsilon_{k}^{2}-m^{2}}\,, &  & \varepsilon_{k}=\frac{m_{J/\psi}^{2}+m^{2}-M^{2}}{2m_{J/\psi}}\,,\label{eq:kinematics}
\end{align}
where $m$~is the mass of the $x$ meson, $m_{J/\psi}$~and $m_{p}$~are
the masses of a $J/\psi$ meson and a proton, respectively, and $\hbar=c=1$.
Since we consider the $p\bar{p}$ invariant mass region \mbox{$M-2m_{p}\ll m_{p}$},
the proton and antiproton are nonrelativistic in their center-of-mass
frame, while $\varepsilon_{k}$~is about $\unit[1]{GeV}$.

The spin-1 wave function of the $p\bar{p}$ pair in the center-of-mass
frame has the form~\cite{dmitriev2014isoscalar}
\begin{equation}
\bm{\psi}_{\lambda}^{I}=\bm{\mathrm{e}}_{\lambda}u_{1}^{I}(0)+\frac{u_{2}^{I}(0)}{\sqrt{2}}\left[\bm{\mathrm{e}}_{\lambda}-3\hat{\bm{p}}(\bm{\mathrm{e}}_{\lambda}\cdot\hat{\bm{p}})\right],
\end{equation}
where $\hat{\bm{p}}=\bm{p}/p$, $\bm{\mathrm{e}}_{\lambda}$~is the
polarization vector of the spin-1 $p\bar{p}$ pair,
\begin{equation}
\sum_{\lambda=1}^{3}\mathrm{e}_{\lambda}^{i}\mathrm{e}_{\lambda}^{j*}=\delta_{ij}\,,
\end{equation}
$u_{1}^{I}(r)$ and $u_{2}^{I}(r)$~are the components of two independent
solutions of the coupled-channels radial Schrödinger equations
\begin{align}
 & \frac{p_{r}^{2}}{m_{p}}\chi_{n}+{\cal V}\chi_{n}=2E\chi_{n}\,,\nonumber \\
 & {\cal V}=\begin{pmatrix}V_{S}^{I} & -2\sqrt{2}\,V_{T}^{I}\\
-2\sqrt{2}\,V_{T}^{I} & \quad V_{D}^{I}-2V_{T}^{I}+{\displaystyle \frac{6}{m_{p}r^{2}}}
\end{pmatrix},\qquad\chi_{n}=\begin{pmatrix}u_{n}^{I}\\
w_{n}^{I}
\end{pmatrix}.
\end{align}
Here $E=p^{2}/2m_{p}$, $V_{S}^{I}$~and $V_{D}^{I}$~are the $N\bar{N}$
potentials in $S$- and $D$-wave channels, and $V_{T}^{I}$~is the
tensor potential. Two independent regular solutions of these equations
are determined by their asymptotic forms at large distances~\cite{dmitriev2014isoscalar}
\begin{align}
 & u_{1}^{I}(r)=\frac{1}{2ipr}\Big[S_{11}^{I}\,e^{ipr}-e^{-ipr}\Big], &  & u_{2}^{I}(r)=\frac{1}{2ipr}S_{21}^{I}\,e^{ipr},\nonumber \\
 & w_{1}^{I}(r)=-\frac{1}{2ipr}S_{12}^{I}\,e^{ipr}, &  & w_{2}^{I}(r)=\frac{1}{2ipr}\Big[-S_{22}^{I}e^{ipr}+e^{-ipr}\Big],
\end{align}
where $S_{ij}^{I}$ are some functions of energy.

The Lorentz transformation for the spin-1 wave function of the $p\bar{p}$
pair can be written~as
\begin{equation}
\tilde{\bm{\psi}}_{\lambda}^{I}=\bm{\psi}_{\lambda}^{I}+\left(\gamma-1\right)\hat{\bm{k}}(\bm{\psi}_{\lambda}^{I}\cdot\hat{\bm{k}})\,,
\end{equation}
where $\tilde{\bm{\psi}}_{\lambda}^{I}$~is the wave function in
the $J/\psi$ rest frame, $\hat{\bm{k}}=\bm{k}/k$, and $\gamma$~is
the $\gamma$-factor of the $p\bar{p}$ center-of-mass frame. The
component collinear to $\bm{k}$ does not contribute to the amplitude
of the decay under consideration because the amplitude is transverse
to $\bm{k}$. As a result, the dimensionless amplitude of the decay
with the corresponding isospin of the $p\bar{p}$ pair can be written~as
\begin{equation}
T_{\lambda\lambda'}^{I}=\frac{\mathcal{G}_{I}}{m_{J/\psi}}\bm{\psi}_{\lambda}^{I}\left[\bm{k}\times\bm{\epsilon}_{\lambda'}\right].\label{eq:amplitude}
\end{equation}
Here $\mathcal{G}_{I}$~is an energy-independent dimensionless constant,
$\bm{\epsilon}_{\lambda'}$~is the polarization vector of~$J/\psi$,
\begin{equation}
\sum_{\lambda'=1}^{2}\epsilon_{\lambda'}^{i}\epsilon_{\lambda'}^{j*}=\delta_{ij}-n^{i}n^{j},
\end{equation}
where $\bm{n}$~is the unit vector collinear to the momentum of electrons
in the beam.

The decay rate of the process $J/\psi\to p\bar{p}x$ can be written
in terms of the dimensionless amplitude $T_{\lambda\lambda'}^{I}$
as (see,~e.g.,~\cite{Sibirtsev2005})
\begin{equation}
\frac{d\Gamma}{dMd\Omega_{p}d\Omega_{k}}=\frac{pk}{2^{9}\pi^{5}m_{J/\psi}^{2}}\left|T_{\lambda\lambda'}^{I}\right|^{2},\label{eq:gamma}
\end{equation}
where $\Omega_{p}$~is the proton solid angle in the $p\bar{p}$
center-of-mass frame and $\Omega_{k}$~is the solid angle of the
$x$ meson in the $J/\psi$ rest frame.

Substituting the amplitude~\eqref{eq:amplitude} in Eq.~\eqref{eq:gamma}
and averaging over the spin states, we obtain the $p\bar{p}$ invariant
mass and angular distribution for the decay rate
\begin{multline}
\frac{d\Gamma}{dMd\Omega_{p}d\Omega_{k}}=\frac{\mathcal{G}_{I}^{2}pk^{3}}{2^{11}\pi^{5}m_{J/\psi}^{4}}\left\{ \left|u_{1}^{I}(0)+{\textstyle \frac{1}{\sqrt{2}}}u_{2}^{I}(0)\right|^{2}+\left|u_{1}^{I}(0)-\sqrt{2}u_{2}^{I}(0)\right|^{2}(\bm{n}\cdot\hat{\bm{k}})^{2}\right.\\
\left.{}+\frac{3}{2}\left[\left|u_{2}^{I}(0)\right|^{2}-2\sqrt{2}\re{\left(u_{1}^{I}(0)u_{2}^{I*}(0)\right)}\right]\left[(\bm{n}\cdot\hat{\bm{p}})^{2}-2(\bm{n}\cdot\hat{\bm{k}})(\bm{n}\cdot\hat{\bm{p}})(\hat{\bm{p}}\cdot\hat{\bm{k}})\right]\right\} .\label{eq:distribution}
\end{multline}
The invariant mass distribution can be obtained by integrating Eq.~\eqref{eq:distribution}
over the solid angles $\Omega_{p}$ and~$\Omega_{k}$:
\begin{equation}
\frac{d\Gamma}{dM}=\frac{\mathcal{G}_{I}^{2}pk^{3}}{2^{5}\thinspace3\pi^{3}m_{J/\psi}^{4}}\left(\left|u_{1}^{I}(0)\right|^{2}+\left|u_{2}^{I}(0)\right|^{2}\right).\label{eq:MassSpectrum}
\end{equation}
The sum in the brackets is the so-called enhancement factor which
equals to unity if the $p\bar{p}$ final-state interaction is turned
off.

More information about the properties of $N\bar{N}$ interaction can
be extracted from the angular distributions. Integrating Eq.~\eqref{eq:distribution}
over $\Omega_{p}$ we obtain
\begin{equation}
\frac{d\Gamma}{dMd\Omega_{k}}=\frac{\mathcal{G}_{I}^{2}pk^{3}}{2^{9}\pi^{4}m_{J/\psi}^{4}}\left(\left|u_{1}^{I}(0)\right|^{2}+\left|u_{2}^{I}(0)\right|^{2}\right)\left[1+\cos^{2}\vartheta_{k}\right],
\end{equation}
where $\vartheta_{k}$~is the angle between $\bm{n}$ and~$\bm{k}$.
However, the angular part of this distribution does not depend on
the features of the $p\bar{p}$ interaction. The proton angular distribution
in the $p\bar{p}$ center-of-mass frame is more interesting. To obtain
this distribution we integrate Eq.~\eqref{eq:distribution} over
the solid angle~$\Omega_{k}$:
\begin{equation}
\frac{d\Gamma}{dMd\Omega_{p}}=\frac{\mathcal{G}_{I}^{2}pk^{3}}{2^{7}\,3\pi^{4}m_{J/\psi}^{4}}\left(\left|u_{1}^{I}(0)\right|^{2}+\left|u_{2}^{I}(0)\right|^{2}\right)\left[1+\gamma^{I}P_{2}(\cos\vartheta_{p})\right],\label{eq:Pdistrib}
\end{equation}
where $\vartheta_{p}$~is the angle between $\bm{n}$ and $\bm{p}$,
$P_{2}(x)=\frac{3x^{2}-1}{2}$~is the Legendre polynomial, and $\gamma^{I}$~is
the parameter of anisotropy:
\begin{equation}
\gamma^{I}=\frac{1}{4}\frac{\left|u_{2}^{I}(0)\right|^{2}-2\sqrt{2}\re{\left[u_{1}^{I}(0)u_{2}^{I*}(0)\right]}}{\left|u_{1}^{I}(0)\right|^{2}+\left|u_{2}^{I}(0)\right|^{2}}\,.\label{eq:anisotropy}
\end{equation}
Averaging~\eqref{eq:distribution} over the direction of $\bm{n}$
gives the distribution over the angle $\vartheta_{pk}$ between $\bm{p}$
and~$\bm{k}$:
\begin{equation}
\frac{d\Gamma}{dMd\Omega_{pk}}=\frac{\mathcal{G}_{I}^{2}pk^{3}}{2^{7}\,3\pi^{4}m_{J/\psi}^{4}}\left(\left|u_{1}^{I}(0)\right|^{2}+\left|u_{2}^{I}(0)\right|^{2}\right)\left[1-2\gamma^{I}P_{2}(\cos\vartheta_{pk})\right].\label{eq:Ppidistrib}
\end{equation}
Note that this distribution can be written in therms of the same anisotropy
parameter~\eqref{eq:anisotropy}.

The mass spectrum~\eqref{eq:MassSpectrum} and the anisotropy parameter~\eqref{eq:anisotropy}
are sensitive to the tensor part of the $N\bar{N}$ potential and,
therefore, gives the possibility to verify the potential model.

\section{Results and Discussion}

\begin{figure}[h]
\includegraphics[height=5.35cm]{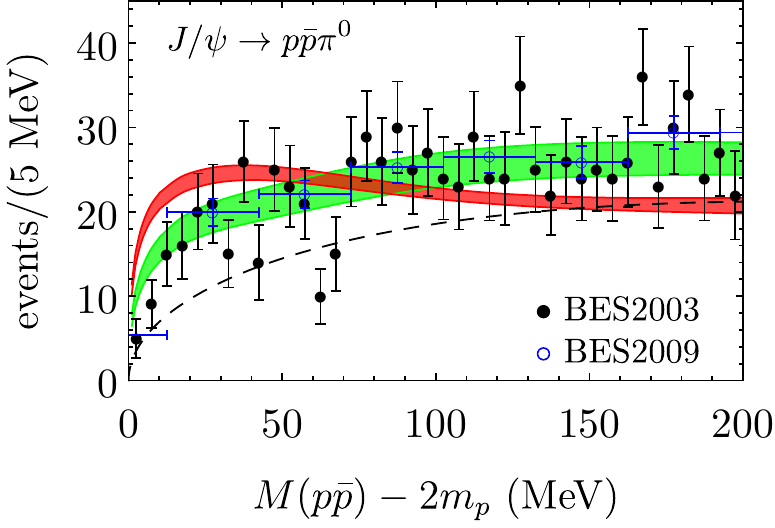}\hfill{}\includegraphics[height=5.35cm]{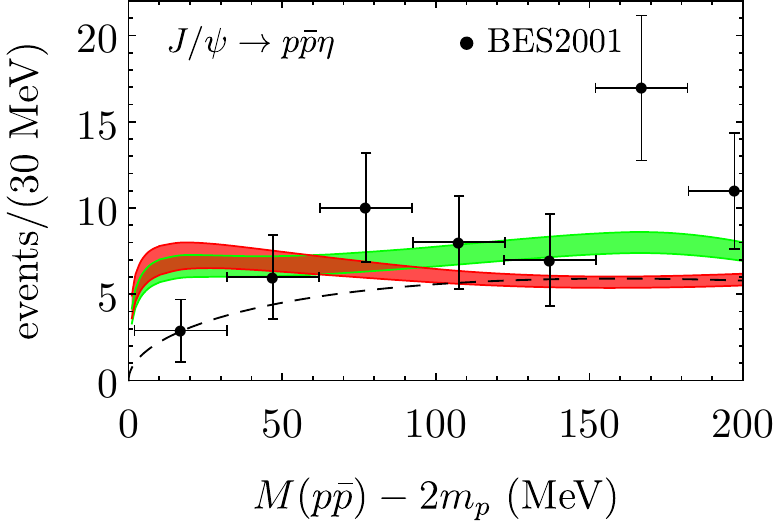}

\caption{\label{fig:JPsidecays}The invariant mass spectra of $J/\psi$ decays
to $p\bar{p}\pi^{0}$ (left) and $p\bar{p}\eta$ (right). The red/dark
band corresponds to the model~\cite{Dmitriev2016} and the green/light
band corresponds to the refitted model. The phase space behavior is
shown by the dashed curve. The experimental data are taken from Refs.~\cite{Bai2003,Ablikim2009,Bai2001}.
The measurement of Ref.~\cite{Bai2003} is adopted for the scale
of the left plot.\hspace*{\fill}}
\end{figure}

In the present work we use the potential model suggested in Ref.~\cite{Dmitriev2016}.
The parameters of this model have been fitted using the $p\bar{p}$
scattering data, the cross section of $N\bar{N}$ pair production
in $e^{+}e^{-}$ annihilation near the threshold, and the ratio of
the electromagnetic form factors of the proton in the timelike region.
By means of this model and Eq.~\eqref{eq:MassSpectrum}, we predict
the $p\bar{p}$ invariant mass spectra in the processes \mbox{$J/\psi\to p\bar{p}\pi^{0}$}
and \mbox{$J/\psi\to p\bar{p}\eta$}. The isospin of the $p\bar{p}$
pair is $I=1$ and $I=0$ for, respectively, a pion and $\eta$ meson
in the final state. The model~\cite{Dmitriev2016} predicts the enhancement
of the decay rates of both processes near the threshold of $p\bar{p}$
pair production (see the red band in Fig.~\ref{fig:JPsidecays}).
The invariant mass spectra predicted by our model are similar to those
predicted in Ref.~\cite{Kang2015} with the use of the chiral model.
Very close to the threshold the enhancement factor turned out to be
slightly overestimated in comparison with the experimental data, as
it is seen from Fig.~\ref{fig:JPsidecays}. We have tried to refit
the parameters of our model in order to achieve a better description
of the invariant mass spectra of the decays considered. The predictions
of the refitted model are shown in Fig.~\ref{fig:JPsidecays} with
the green band. It is seen that the refitted model fits better the
invariant mass spectra of $J/\psi$ decays. However, the discrepancy
in the cross sections of $n\bar{n}$ production in $e^{+}e^{-}$ annihilation
and the charge-exchange process \mbox{$p\bar{p}\to n\bar{n}$} have
slightly increased after refitting.

\begin{figure}
\includegraphics[height=5.35cm]{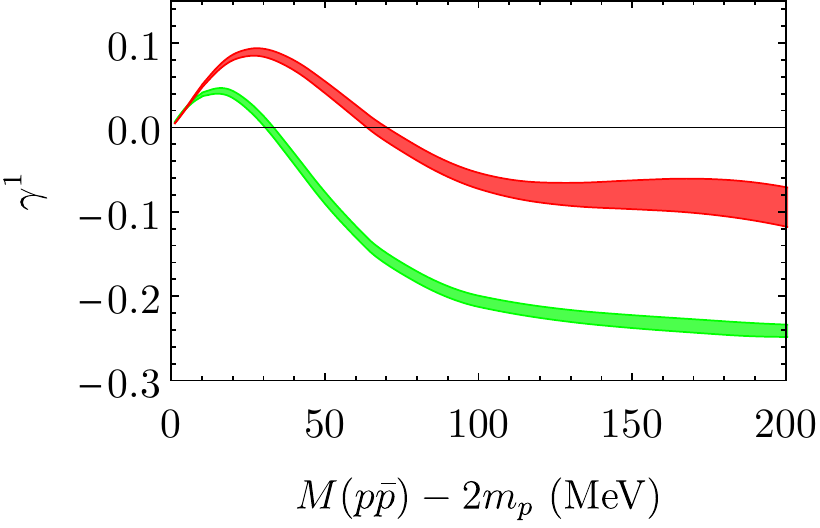}\hfill{}\includegraphics[height=5.35cm]{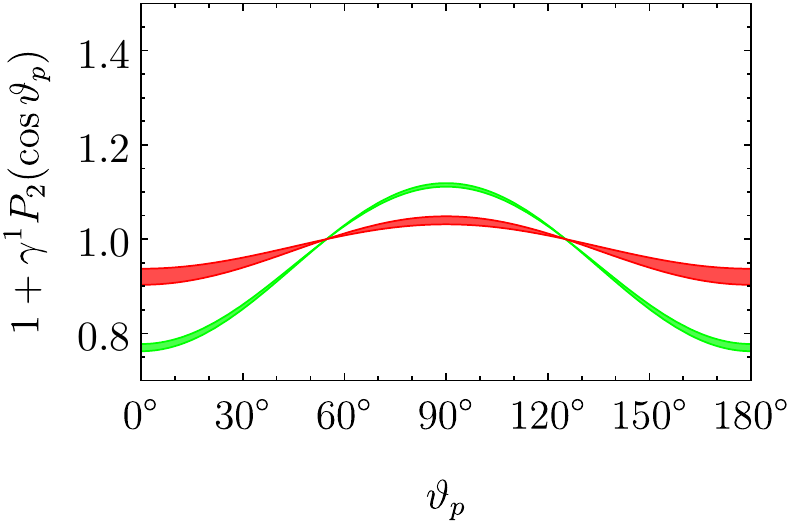}

\includegraphics[height=5.35cm]{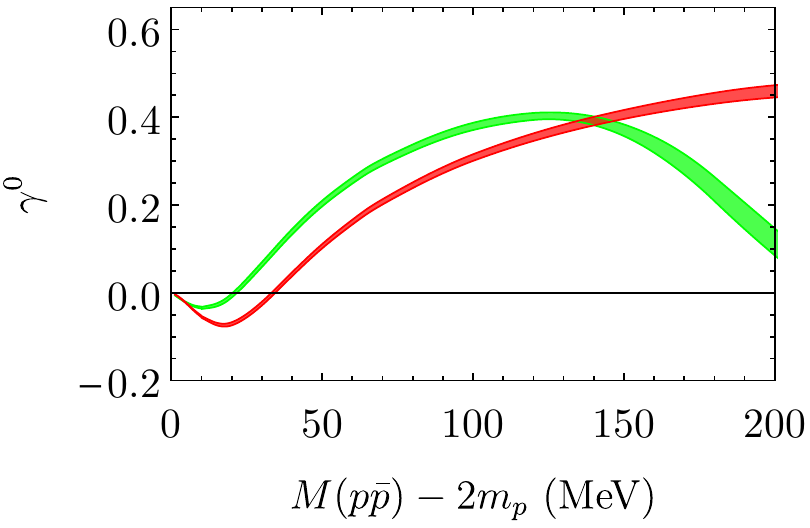}\hfill{}\includegraphics[height=5.35cm]{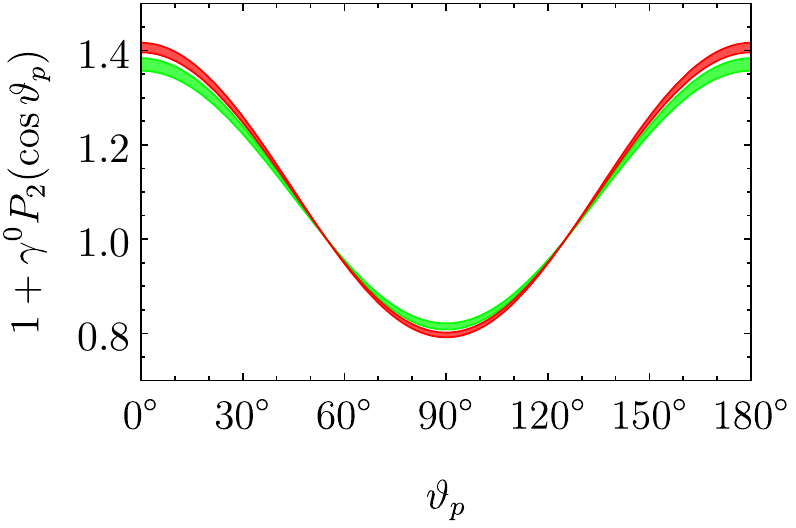}

\caption{\label{fig:anisotropy}The dependence of the anisotropy parameters
$\gamma^{I}$ on  $p\bar{p}$ invariant mass (left) and the distributions
over the angle between the proton momentum and the momentum of the
electrons in the beam at $M-2m_{p}=\unit[150]{MeV}$ (right). The
red/dark band corresponds to the model~\cite{Dmitriev2016} and the
green/light band corresponds to the refitted model.\hspace*{\fill}}
\end{figure}

An important prediction of our model is the angular anisotropy of
the $J/\psi$ decays. This anisotropy is the result of $D$-wave admixture
due to the tensor forces in $N\bar{N}$ interaction. The anisotropy
(see Eqs.~\eqref{eq:Pdistrib} and~\eqref{eq:Ppidistrib}) is characterized
by the parameters $\gamma^{1}$ and $\gamma^{0}$~\eqref{eq:anisotropy}
for the $p\bar{p}\pi^{0}$ and $p\bar{p}\eta$ final states, respectively.
The dependence of the parameters $\gamma^{I}$ on the invariant mass
of the $p\bar{p}$ pair is shown in the left side of Fig.~\ref{fig:anisotropy}.
For $p\bar{p}$ invariant mass about $\unit[100-200]{MeV}$ above
the threshold, significant anisotropy of the angular distributions
is predicted. The distributions over the angle between the proton
momentum and the momentum of the electrons in the beam are shown in
the right side of Fig.~\ref{fig:anisotropy}. Note that the anisotropy
in the distribution over the angle $\vartheta_{pk}$ is expected to
be two times larger than in the distribution over the angle $\vartheta_{p}$
(compare Eqs.~\eqref{eq:Pdistrib} and~\eqref{eq:Ppidistrib}).

There are some data on the angular distributions in the decays \mbox{$J/\psi\to p\bar{p}\pi^{0}$}
~\cite{Ablikim2009} and \mbox{$J/\psi\to p\bar{p}\eta$} ~\cite{Bai2001}.
However, these distributions are obtained by integration over the
whole $p\bar{p}$ invariant mass region. Unfortunately, our predictions
are valid only in the narrow energy region above the $p\bar{p}$ threshold.
Therefore, we cannot compare the predictions with the available experimental
data. The measurements of the angular distributions at $p\bar{p}$
invariant mass close to the $p\bar{p}$ threshold would be very helpful.
Such measurements would provide another possibility to verify the
available models of $N\bar{N}$ interaction in the low-energy region.

\begin{figure}
\includegraphics[height=5.35cm]{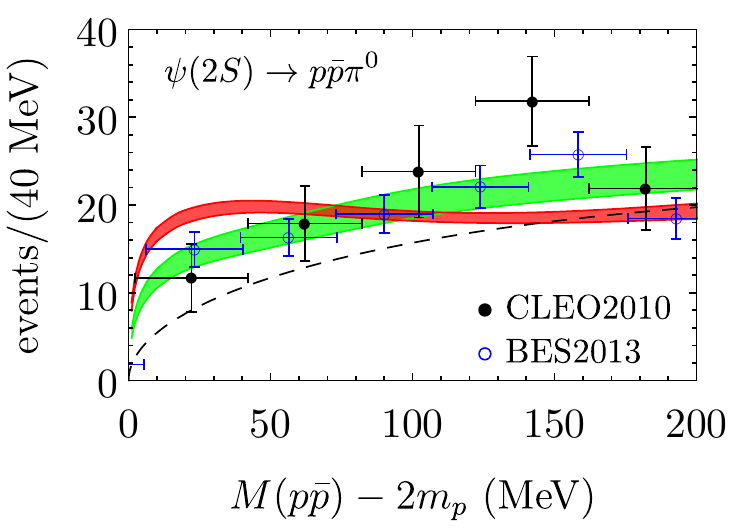}\hfill{}\includegraphics[height=5.35cm]{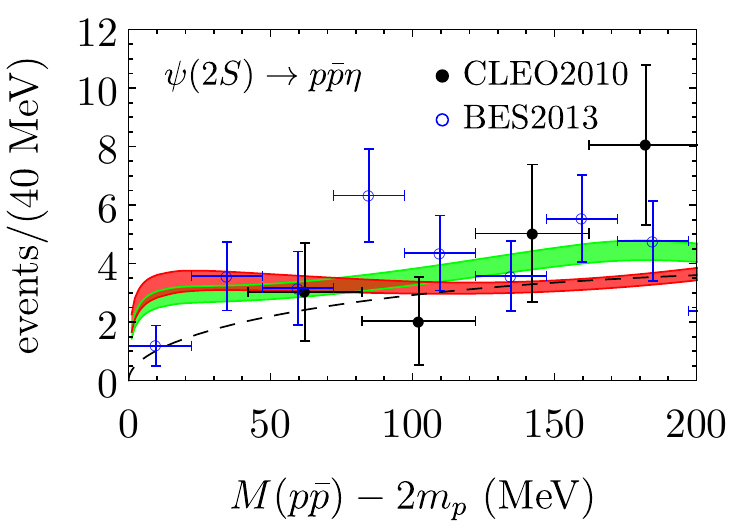}

\caption{\label{fig:PsiPrime}The invariant mass spectra for the decays $\psi(2S)\to p\bar{p}\pi^{0}$
(left) and $\psi(2S)\to p\bar{p}\eta$ (right). The red/dark band
corresponds to the model~\cite{Dmitriev2016} and the green/light
band corresponds to the refitted model. The phase space behavior is
shown by the dashed curve. The experimental data are taken from Refs.~\cite{Alexander2010,Ablikim2013a,Ablikim2013}.
The measurement of Ref.~\cite{Alexander2010} is adopted for the
scale of both plots.\hspace*{\fill}}
\end{figure}

The formulas written above are also valid for the decays \mbox{$\psi(2S)\to p\bar{p}\pi^{0}$}
and \mbox{$\psi(2S)\to p\bar{p}\eta$} with the replacement of $m_{J/\psi}$
by the mass of $\psi(2S)$. The invariant mass spectra for these decays
are shown in Fig.~\ref{fig:PsiPrime}. The angular distributions
for these processes are the same as for the decays of $J/\psi$ because
they depend only on the invariant mass of the $p\bar{p}$ pair.

\section{Conclusions}

Using the model proposed in Ref.~\cite{Dmitriev2016}, we have calculated
the effects of $p\bar{p}$ final-state interaction in the decays \mbox{$J/\psi\to p\bar{p}\pi^{0}(\eta)$}
and \mbox{$\psi(2S)\to p\bar{p}\pi^{0}(\eta)$}. Our results for
the $p\bar{p}$ invariant mass spectra close to the $p\bar{p}$ threshold
are in agreement with the available experimental data. The tensor
forces in the $p\bar{p}$ interaction result in the anisotropy of
the angular distributions. The anisotropy in the decay \mbox{$J/\psi\to p\bar{p}\pi^{0}$}
and especially in the \mbox{$J/\psi\to p\bar{p}\eta$} decay are
large enough to be measured. The observation of such anisotropy close
to the $p\bar{p}$ threshold would allow one to refine the model of
$N\bar{N}$ interaction.

\bibliographystyle{/home/sergey/Documents/BINP/BibTeX/apsrev4-1}
\bibliography{/home/sergey/Documents/BINP/BibTeX/library}

\end{document}